# Influence of interphase boundary coherency in high-entropy alloys on their hydrogen storage performance

Shivam Dangwal[1,2] and Kaveh Edalati[1,2,]*

[1] WPI, International Institute for Carbon-Neutral Energy Research (WPI-I2CNER), Kyushu University, Fukuoka 819-0395, Japan
[2] Department of Automotive Science, Graduate School of Integrated Frontier Sciences, Kyushu University, Fukuoka 819-0395, Japan

High-entropy alloys (HEAs) have potential for storing hydrogen reversibly at room temperature due to their tunable thermodynamics; however, they usually suffer from the issue of difficult activation. This study shows that while interphase boundaries are effective in activating some HEAs, some other dual-phase HEAs still require extra high-temperature activation. To understand why interphase boundaries are not always effective for activation, microstructural features and hydrogen storage performance of six HEAs with dual phases are compared. Detailed analysis confirms that interphase boundaries are effective for hydrogen absorption without the need for activation treatment, provided that: (i) their fraction is high enough, and (ii) they are not coherent. These findings are discussed in terms of free volume and boundary energy. Coherent interphase boundaries are associated with lower free volume and thus do not act as fast hydrogen diffusion paths. Moreover, the boundary energy of coherent boundaries is lower than incoherent boundaries, making them less favorable sites for heterogeneous hydride nucleation. This research thus suggests that the introduction of incoherent interphase boundaries with a proper fraction is a solution for activating hydrogen storage materials.

***Keywords:*** high-entropy hydrides; metal hydrides; hydrogen fuel; grain orientation, multi-principal element alloys (MPEAs)

*Correspondence (Email: kaveh.edalati@kyudai.jp; Phone: +81-92-802-6744)

# 1. Introduction

The global energy scene is transforming as countries attempt to lower their carbon emissions. Countries such as Japan, India, France and China are considering hydrogen as a part of their goal to lower carbon emissions [1]. Hydrogen can play a crucial role in lowering carbon emissions because it is a zero-emission energy carrier. The combustion of hydrogen produces water, which is non-polluting. An efficient utilization of hydrogen can be seen in fuel cell vehicles, where the electrochemical reaction of hydrogen and oxygen produces water and electricity [1,2]. Although a fuel cell with 100% efficiency is thermodynamically impossible, fuel cells still provide more efficient hydrogen consumption than combustion [3]. The gravimetric energy density of hydrogen is 120 MJ/kg, which is greater than other fuels [1,2,4], but its volumetric energy density is not high under standard temperature and pressure [4]. The volumetric density can be enhanced if hydrogen is stored in the metal hydride form, a solid-state storage technique [5]. This technique not only increases volumetric density but also provides safety by lowering the operation pressure [2,5]. Researchers in the field of solid-state hydrogen storage have progressed creatively through both experimental investigations and fundamental studies at the atomic and quantum chemistry levels.

Theoretical studies have been employed to calculate the hydrogen binding energy and hydrogen interaction with metals and alloys [6]. For room-temperature hydrogen storage, binding energy for alloys should lie close to -0.1 eV [6,7] or the formation enthalpy of hydride should be between -25 to -39 kJ/mol [8]. Beyond traditional hydrides, theoretical studies were also developed for non-conventional materials such as pyridine-Li complex [9], heterocyclic frameworks with Ag(I)/Au(I) [10], Au(I) complex [11], metal-organic frameworks [12,13], hetero-doped carbon materials [14], and coal-derived nanocarbons [15]. All these studies suggest that precise control over atomic structure and surface characteristics is critical for enhancing the performance of hydrogen storage materials. Despite significant progress on complex systems, conventional metal hydrides are still the most popular hydrogen storage materials.

There are various metals, like magnesium [16] and lithium [17,18], as well as intermetallics, like $LaNi_5$ [19] and TiFe [20,21], that show hydrogen storage capacity. However, these metals and intermetallics have some operational issues. For example, magnesium [16] shows slow absorption and requires high temperatures close to 763 K for desorption. Lithium [17,18] forms a very stable hydride and requires a temperature over 976 K for hydrogen desorption. $LaNi_5$

[19] is expensive, and TiFe [20,21] suffers from activation issues. In fact, most metals and intermetallics either suffer from activation issues or do not thermodynamically work at room temperature. The issue of activation can be solved by different methods, such as high-temperature treatment [2,5,20,22] and mechanical treatment (ball milling [23,24] and severe plastic deformation [25,26]) to introduce grain boundaries. However, these solutions usually add an additional step to the processing route of materials [27]. A traditional strategy of controlling boundaries by alloying can avoid this additional processing step issue, as it has been used successfully for different systems, such as core-shell morphologies [28], dual phases [29] or precipitates [30]. Moreover, some studies considered severe alloying and the formation of high-entropy alloys (HEAs) as a solution to both thermodynamics and activation issues for solid-state hydrogen storage and nickel-metal hydride batteries [31].

HEAs are made by mixing five or more elements to generate an entropy over $1.5R$ ($R$: gas constant) [28]. These alloys are primarily considered solid solutions, although recent definitions also include ordered phases such as intermetallics and ceramics as HEAs [32]. The presence of multiple metals in HEAs associates them with features such as sluggish diffusion, high entropy, severe lattice distortion and cocktail effect [32,33]. These unique features distinguish them from traditional metals and alloys, leading to versatile applications as catalytic materials [34,35], refractory materials [36], biomaterials [37], structural materials [38,39,40,41], etc. Additionally, HEAs have the thermodynamic potential to work at room temperature for solid-state hydrogen storage [8,42]. Although some studies have shown reversible storage capacity without any activation in some HEAs [43-45], the majority of HEAs suffer from the issue of activation. A few examples of some entropy-stabilized alloys that require activation are Mg-Ti-V-Zr-Nb [46], TiVZrHfNb [47], TiVZrNbTa [48], TiZrHfMoNb [49], TiZrNbHfTa [50], TiZrFeNi [51] and ZrNbFeCo [52]. A recent study showed that a solution to the activation issue of HEAs is the use of dual-phase materials with interphase boundaries [45]. Interphase boundaries can function as a path to transfer hydrogen atoms to bulk from surface layers as well as for heterogenous hydride nucleation, an action similar to high-angle grain boundaries [45,53-56]. However, the current authors found that some dual-phase HEAs still require activation. The key question is when interphase boundaries are effective for activation and when they are not.

One factor that may influence the effectiveness of interphase boundaries is the orientation of the grains, their misorientation and the coherency of interphase boundaries. Hydrogen diffusion is known to depend on grain orientation due to the differences in surface energies of different atomic planes [57]. Additionally, hydrogen mobility through a material is influenced by the misorientation between the neighboring grains [58]. Although prior studies investigated the role of grain orientation and misorientation on hydrogen embrittlement [57,58,59] and catalytic hydrogen evolution reaction [35], no study directly addresses the influence of interphase misorientation and coherency on hydrogen absorption in HEAs. Therefore, exploring how grain orientation of two phases and interphase coherency affect hydrogen absorption, particularly in dual-phase HEAs, is an unexplored research direction.

The objective of this study is thus to answer the above question of why some dual-phase HEAs with interphase boundaries still require activation for hydrogen absorption. To investigate this, six different HEAs are synthesized by considering elemental configuration, valence electron concentration and C14 phase stability. Two HEAs ($Ti_{0.5}Zr_{1.5}CrMnFeNi$ and $TiZrCrMnFeNi$) have $AB_2$-type configuration and four HEAs ($TiV_{1.5}ZrCr_{0.5}MnFeNi$, $TiV_2ZrCrMnFeNi$, $Ti_2VCrMnFe$ and $TiV_{1.5}Zr_{1.5}CrMnFeNi$) have AB-type configuration, where A stands for elements that have a high hydrogen affinity and B stands for those with a low hydrogen affinity. In addition to AB- or $AB_2$-type configurations, the compositions are selected to have a valence electron concentration close to 6.4, which was reported to be appropriate for room-temperature hydrogen storage [8,31]. All six HEAs have a major C14 Laves phase, due to the co-presence of zirconium with titanium, and a minor fraction of a body-centered cubic (BCC) structure as the secondary phase, in agreement with the authors' earlier studies on some of these alloys [42-45]. This study reports for the first time that incoherent interphase boundaries with appropriate volume fractions are effective in activating HEAs, whereas coherent interphase boundaries are not effective. This issue is discussed by considering excess free volume and boundary energy differences between the coherent and incoherent boundaries [60-65].

## 2. Experimental Procedures

Ingots of six HEAs were synthesized by vacuum arc melting. Metals (>99% purity) were utilized to produce HEAs in a water-cooled copper mold under argon atmosphere, as attempted earlier [42-45]. The homogeneous composition of HEAs was ensured by flipping and remelting each HEA seven times during the process of arc melting.

After arc melting, an X-ray diffractometer (XRD) was used for phase analysis. X-ray light was generated by Cu Kα, and crystallographic phases and their lattice parameters were determined for alloyed elements through Rietveld refinement using PDXL software (used JCPDS card numbers for C14 and BCC were 74-3259 and 77-8144, respectively).

The microstructures of HEAs were studied through a scanning electron microscope (SEM). To prepare SEM samples, a disc from the center of the ingots was cut using electrical discharge machining. To achieve mirror-like surfaces on these disc-shaped samples, they were first ground with sandpapers (grids: 400, 800, 1200 and 2000), then fine-ground with diamond suspensions (sizes: 9 and 3 µm) and a buff, and polished to the final form with 60 nm colloidal silica. The microstructure of materials was examined via electron backscatter diffraction (EBSD) using a voltage of 15 kV. Post-EBSD analysis was conducted by employing the MTEX toolbox in MATLAB [66]. Grain boundary smoothing was conducted by giving the mean orientation of neighboring grains to unidentified regions by utilizing the grain property meanOrientation [66]. Euler angles obtained from EBSD were converted to Miller indices using Eq. (1) and (2) for the BCC phase and to Miller-Bravais indices using Eq. (3) and (4) for the C14 Laves phase [67,68].

$$\begin{bmatrix} h \\ k \\ l \end{bmatrix} = \begin{bmatrix} \sin \varphi_2 \sin \Phi \\ \cos \varphi_2 \sin \Phi \\ \cos \Phi \end{bmatrix} \quad (1)$$

$$\begin{bmatrix} u \\ v \\ w \end{bmatrix} = \begin{bmatrix} \cos \varphi_1 \cos \varphi_2 - \sin \varphi_1 \sin \varphi_2 \cos \Phi \\ -\cos \varphi_1 \sin \varphi_2 - \sin \varphi_1 \cos \varphi_2 \cos \Phi \\ \sin \varphi_1 \sin \Phi \end{bmatrix} \quad (2)$$

$$\begin{bmatrix} h \\ k \\ i \\ l \end{bmatrix} = \begin{bmatrix} \frac{\sqrt{3}}{2} & -\frac{1}{2} & 0 \\ 0 & 1 & 0 \\ -\frac{\sqrt{3}}{2} & -\frac{1}{2} & 0 \\ 0 & 0 & c/a \end{bmatrix} \begin{bmatrix} \sin \varphi_2 \sin \Phi \\ \cos \varphi_2 \sin \Phi \\ \cos \Phi \end{bmatrix} \quad (3)$$

$$\begin{bmatrix} u \\ v \\ t \\ w \end{bmatrix} = \begin{bmatrix} \frac{2}{3} & -\frac{1}{3} & 0 \\ 0 & \frac{2}{3} & 0 \\ -\frac{2}{3} & -\frac{1}{3} & 0 \\ 0 & 0 & c/a \end{bmatrix} \begin{bmatrix} \cos\varphi_1 \cos\varphi_2 - \sin\varphi_1 \sin\varphi_2 \cos\Phi \\ -\cos\varphi_1 \sin\varphi_2 - \sin\varphi_1 \cos\varphi_2 \cos\Phi \\ \sin\varphi_1 \sin\Phi \end{bmatrix} \quad (4)$$

where $(h\ k\ l)$ and $[u\ v\ w]$ are Miller indices of planes and directions of the BCC phase, respectively, $(h\ k\ i\ l)$ and $[u\ v\ t\ w]$ are Miller-Bravais indices of planes and directions of the C14 Laves phase, respectively, $(h\ k\ l)[u\ v\ w]$ represents crystal orientation with $(h\ k\ l)$ in the normal direction and $[u\ v\ w]$ in the rolling direction, $\varphi_1$, $\Phi$ and $\varphi_1$ are Euler angles representing the rotation of the crystal along the normal direction, rolling direction and again normal direction, respectively, and $a$ and $c$ are the lattice parameters for C14.

The nanostructural characterizations were conducted by a transmission electron microscope (TEM). TEM foils were made by crushing the ingots in alcohol, dispersing them on a carbon grid, and examining them under 200 kV by high-resolution TEM images.

Hydrogen storage performance was analyzed in a Sievert-type machine using pressure-composition-temperature (PCT) isotherms at 303 K. Sample preparation for PCT measurements was done by crushing the ingots and passing them through a 75 µm sieve. The hydrogenation test was first carried out at 303 K with no activation. When an alloy did not show hydrogenation at room temperature, it was activated in vacuum at 673 K for 60 min and subsequently re-examined by PCT at 303 K.

## 3. Results
### 3.1. Hydrogen Storage Performance

Hydrogen storage results at 303 K using PCT isotherms and without activation treatment are illustrated in Fig. 1(a) for $Ti_{0.5}Zr_{1.5}CrMnFeNi$, $TiZrCrMnFeNi$, $TiV_{1.5}ZrCr_{0.5}MnFeNi$, and $TiV_2ZrCrMnFeNi$, and in Fig. 1(b) for $Ti_2VCrMnFe$ and $TiV_{1.5}Zr_{1.5}CrMnFeNi$. Four HEAs in Fig. 1(a) show reversible hydrogen storage capacity without any activation treatment in their first cycle, and hence, they are active HEAs. Fig. 1(b) shows PCT isotherms for the two remaining HEAs, $Ti_2VCrMnFe$ and $TiV_{1.5}Zr_{1.5}CrMnFeNi$, at 303 K without any activation process. These two HEAs did not absorb hydrogen at 303 K without any activation process. Therefore, they were activated at 673 K for 1 h and examined again by PCT at 303 K, as illustrated in Fig. 1(c). Both alloys, after activation, are hydrogenated at ambient temperature, suggesting the drawback of these two alloys is their difficult activation (no thermodynamic stability drawback), as usually expected for various room-temperature hydrogen storage materials, including conventional materials [69-71] and high-entropy alloys [72-74]. The reason for the activation by high-temperature treatment has been a matter of argument for decades, but it can be due to the formation of active catalysts on the surface or partial reduction of the oxide layer on the surface [75]. In the next sessions, the features of these alloys are compared to understand the reasons for the activity of four of them and the inactivity of two of them.

**3.2. Phase Structures**

XRD profiles of HEAs are shown in Fig. 2. The XRD profiles reveal the existence of major C14 Laves phase (*P63/mmc*) and the secondary phase BCC ($Im\bar{3}m$), as in all HEAs. It should be noted that no face-centered cubic (FCC) phase was detected before or after hydrogenation. The presence of C14 is promising because its ability to absorb/desorb hydrogen has been frequently reported in the literature for conventional alloys [69-71] and HEAs [72-74,76]. The details of crystal structures and lattice parameters examined by the Rietveld method are shown in Table 1. Moreover, the peak corresponding to the (110) plane of BCC was compared with the nearest peak of the C14 Laves phase, and their *d*-spacing difference was calculated and summarized in Table 1, because the difference in *d*-spacing is an important factor in determining the coherency at interphase boundaries. Phase fraction for the BCC phase is the highest for TiV$_2$ZrCrMnFeNi with 16 wt%, and the lowest for the TiV$_{1.5}$Zr$_{1.5}$CrMnFeNi with less than 1 wt%. For HEAs TiV$_{1.5}$Zr$_{1.5}$CrMnFeNi, it is observed that the most intensive BCC peak, corresponding to (110), coincides with the peak of C14. This peak coincidence leads to a very low *d*-spacing difference of less than 0.03%, when (112)$_{C14}$ // (110)$_{BCC}$ for Ti$_2$VCrMnFe. Moreover, the *d*-spacing difference for TiV$_2$ZrCrMnFeNi is also small, around 0.6%. Small *d*-spacing difference between two phases suggests the possibility of the coherency of interphase boundaries. To understand more about coherency, grain orientation was studied using EBSD.

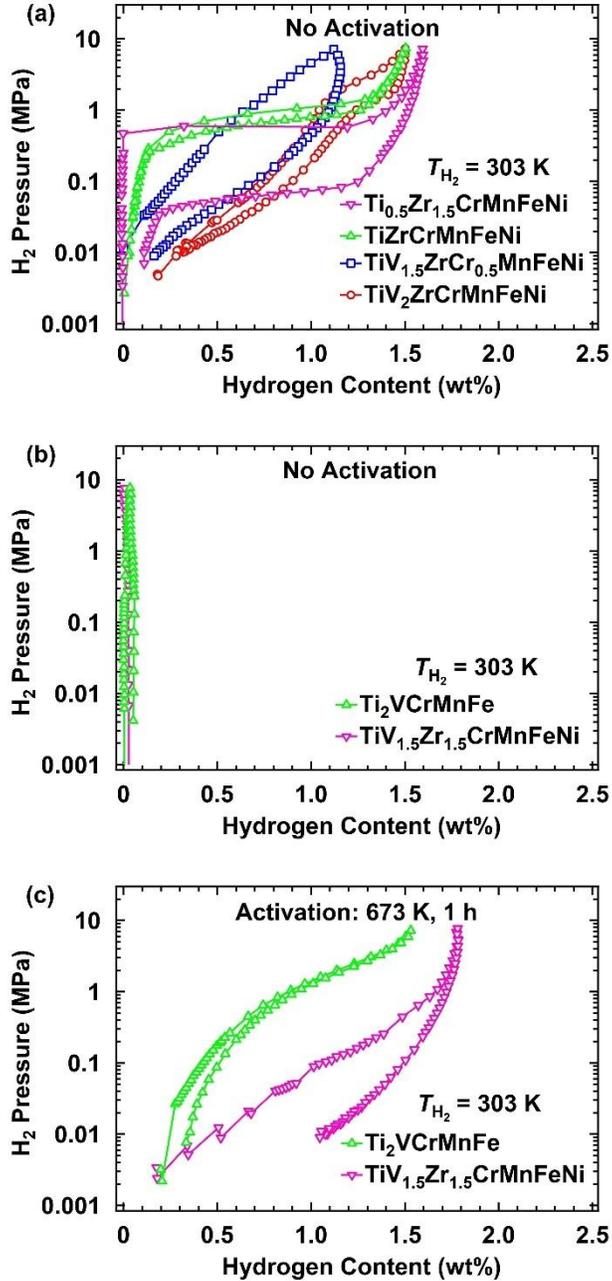

Figure 1. Reversible hydrogen storage at 303 K in high-entropy alloys without or with activation treatment. Pressure-composition-temperature isotherms at 303 K for (a) $Ti_{0.5}Zr_{1.5}CrMnFeNi$, $TiZrCrMnFeNi$, $TiV_{1.5}ZrCr_{0.5}MnFeNi$ and $TiV_2ZrCrMnFeNi$ without any activation treatment, (b) $Ti_2VCrMnFe$ and $TiV_{1.5}Zr_{1.5}CrMnFeNi$ without any activation treatment, and (c) $Ti_2VCrMnFe$ and $TiV_{1.5}Zr_{1.5}CrMnFeNi$ following activation at 673 K for 60 min.

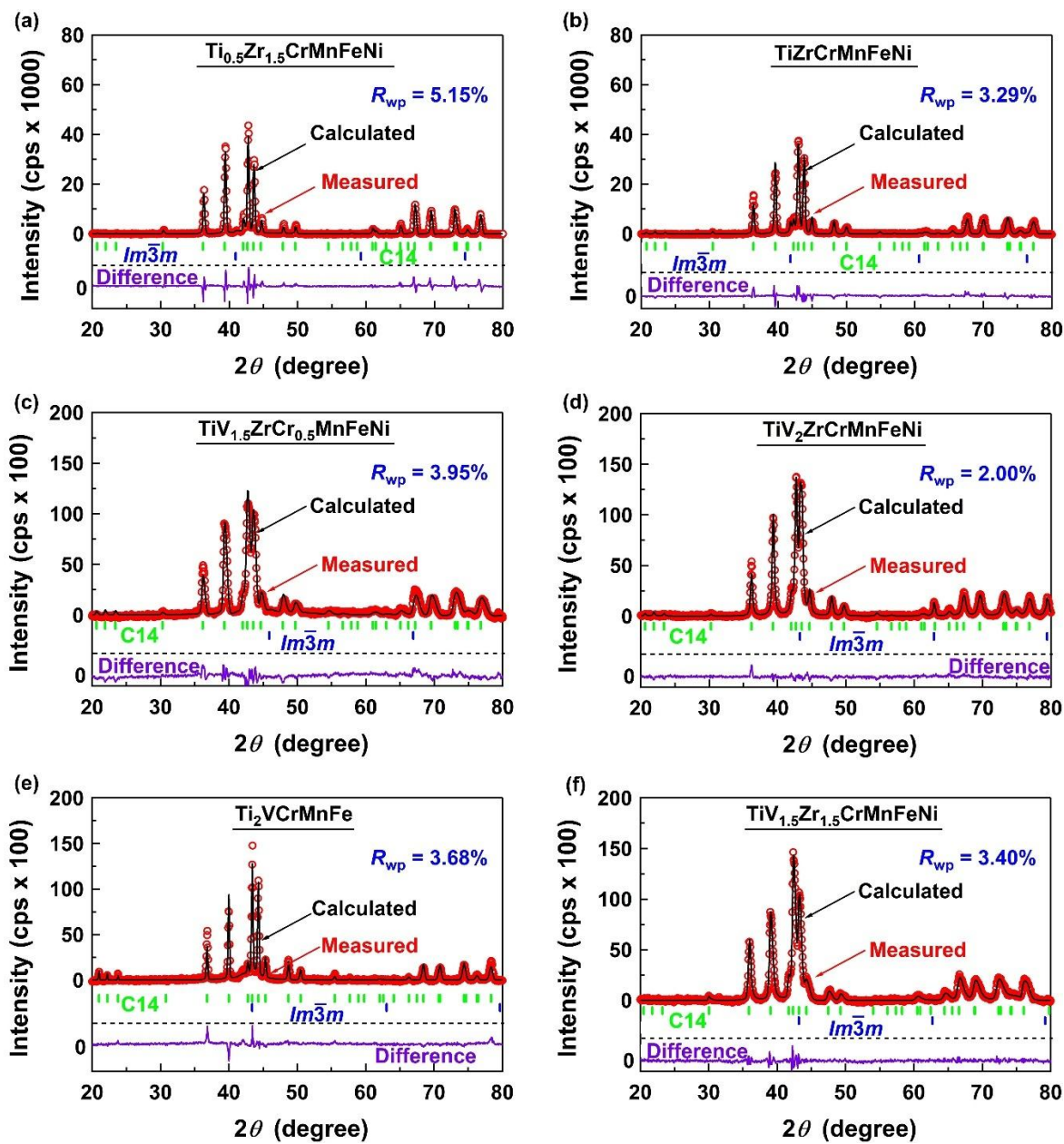

Figure 2. High-entropy alloys with C14 Laves and BCC phases. XRD profiles and corresponding Rietveld analyses ($R_{wp}$: weighted profile $R$-factor) for (a) $Ti_{0.5}Zr_{1.5}CrMnFeNi$, (b) $TiZrCrMnFeNi$, (c) $TiV_{1.5}ZrCr_{0.5}MnFeNi$, (d) $TiV_2ZrCrMnFeNi$, (e) $Ti_2VCrMnFe$ and (f) $TiV_{1.5}Zr_{1.5}CrMnFeNi$.

Table 1. Lattice parameters, BCC phase fractions, and *d*-spacing difference between (110) plane of BCC and closer atomic planes of C14, such as (112), (201), (004) and (200), and activity for hydrogen storage for six high-entropy alloys used in current investigation.

| High-entropy alloy | BCC (wt%) | Lattice parameters of C14 (Å) | Lattice parameter of BCC (Å) | C14/BCC *d*-spacing difference (%) | Activity |
|---|---|---|---|---|---|
| $Ti_{0.5}Zr_{1.5}CrMnFeNi$ | 2 | $a = 4.97, c = 8.13$ | $a = 3.12$ | 2.501 | Active |
| $TiZrCrMnFeNi$ | 2 | $a = 4.93, c = 8.06$ | $a = 3.05$ | 1.049 | Active |
| $TiV_{1.5}ZrCr_{0.5}MnFeNi$ | 1 | $a = 4.96, c = 8.11$ | $a = 2.79$ | 2.579 | Active |
| $TiV_2ZrCrMnFeNi$ | 16 | $a = 4.96, c = 8.10$ | $a = 2.95$ | 0.592 | Active |
| $Ti_2VCrMnFe$ | 7 | $a = 4.88, c = 7.99$ | $a = 2.95$ | 0.03 | Inactive |
| $TiV_{1.5}Zr_{1.5}CrMnFeNi$ | <1 | $a = 5.00, c = 8.18$ | $a = 2.96$ | 0.008 | Inactive |

### 3.3. Microstructure of High-Entropy Alloys Examined by EBSD

EBSD phase and orientation mappings as well as inverse pole figures (IPF) for $AB_2$-type HEAs $Ti_{0.5}Zr_{1.5}CrMnFeNi$ and $TiZrCrMnFeNi$ are illustrated in Fig. 3(a-c) and Fig. 3(d-f), respectively. The co-existence of C14 and BCC is confirmed from the EBSD phase map for both HEAs, which is consistent with XRD results. Different colors of BCC orientation within the same C14 matrix are observed in both HEAs. This confirms that random misorientation between C14 and BCC exists for both HEAs. The IPF images show the spread of different BCC orientations in the HEAs. These EBSD results suggest that there is no preferred coherency between the C14 matrix and the secondary phase BCC.

Fig. 4 illustrates the EBSD phase map, orientation map and IPF images for AB-type high-entropy materials, with $TiV_{1.5}ZrCr_{0.5}MnFeNi$ shown in Fig. 4(a-c) and $TiV_2ZrCrMnFeNi$ illustrated in Fig. 4(d-f). Similar to the results for the $AB_2$-type HEAs, these HEAs also show the random misorientation between C14 and BCC. The IPF images also show the spread of different BCC orientations in both HEAs. Therefore, EBSD results suggest that the C14 and BCC interphases should not be necessarily coherent.

EBSD results are displayed in Fig. 5(a-c) for $Ti_2VCrMnFe$ and in Fig. 5(d-f) for $TiV_{1.5}Zr_{1.5}CrMnFeNi$. EBSD phase map for the HEA $Ti_2VCrMnFe$ in Fig. 5(a) indicates the co-presence of C14 and BCC. The orientation map for the HEA $Ti_2VCrMnFe$ shows the same BCC orientation within the matrix of C14 for all BCC grains. This confirms uniform orientation between C14 and BCC and the possibility for the formation of coherent boundaries, unlike previous HEAs. This is consistent with the small *d*-spacing differences shown for this alloy in Table 1, as a smaller difference in *d*-spacing between C14 and BCC enhances the likelihood of coherency at the interface. The IPF images for BCC in Fig. 5(c) also show a high value for multiples of uniform density (MUD) around 12, which indicates stronger alignment of specific planes to produce coherent interphase boundaries [77]. It is known from the literature that coherent interfaces have lower energy levels, whereas incoherent interfaces have higher energies [78,79], and thus, secondary grains like to form coherent boundaries when *d*-spacing differences allow. EBSD phase map, orientation map and IPF for the HEA TiV$_{1.5}$Zr$_{1.5}$CrMnFeNi in Fig. 5(d, e, f) suggest random misorientation between the C14 and BCC phases in this HEA, although the fraction of BCC is very low. The presence of a low portion of interphase boundaries is not expected to affect the hydrogenation properties of this HEA.

The orientation angle between the C14 and BCC phases is plotted against the frequency as shown in Fig. 6. The plotted curve represents a normalized misorientation distribution function (MDF), where mrd values quantify the relative frequency of specific misorientations compared to a random distribution. Examination of Fig. 6 indicates the following features.

- For the AB$_2$-type HEA Ti$_{0.5}$Zr$_{1.5}$CrMnFeNi and TiZrCrMnFeNi, random orientation is observed with various peaks along a fiber, as shown Fig. 6(a) and Fig. 6(b), respectively. These random orientations enhance the hydrogen diffusion and thus are effective in making materials active for hydrogenation [60,65].
- A major uniform orientation is observed for the HEA TiV$_{1.5}$ZrCr$_{0.5}$MnFeNi as shown in Fig. 6(c). Despite having major uniform misorientation, TiV$_{1.5}$ZrCr$_{0.5}$MnFeNi is an active HEA due to the presence of incoherent interphase boundaries with large *d*-spacing differences. Moreover, the misorientation angle between C14 and BCC for this HEA lies between 30º and 50º, which is appropriate for a higher hydrogenation rate [60]. In line with the Frank-Bilby model, the density of defects at a boundary rises continuously for misorientations up to 36.87° and then gradually decreases after 36.87º [60,80,81]. This high defect density, combined with the incoherency of the interphase boundary, makes TiV$_{1.5}$ZrCr$_{0.5}$MnFeNi an active material for hydrogen absorption.
- Random orientations between the C14 and BCC phases are observed for the AB-type HEA TiV$_2$ZrCrMnFeNi, as shown in Fig. 6(d), proving the activity of the HEA.

- For the HEA Ti$_2$VCrMnFe, high frequency is observed for misorientations more than 50º as shown in Fig. 6(e). It is known from the literature that the hydrogenation rate is lower for misorientation angles between 50° to 58º [60]. This misorientation, combined with the coherency of the interphase boundary, makes Ti$_2$VCrMnFe an inactive material for hydrogen absorption.
- TiV$_{1.5}$Zr$_{1.5}$CrMnFeNi in Fig. 6(f) shows orientation angles in the range where the hydrogenation rate is higher, but the portion of interphase boundaries is not high enough to impact its hydrogenation [54-56].

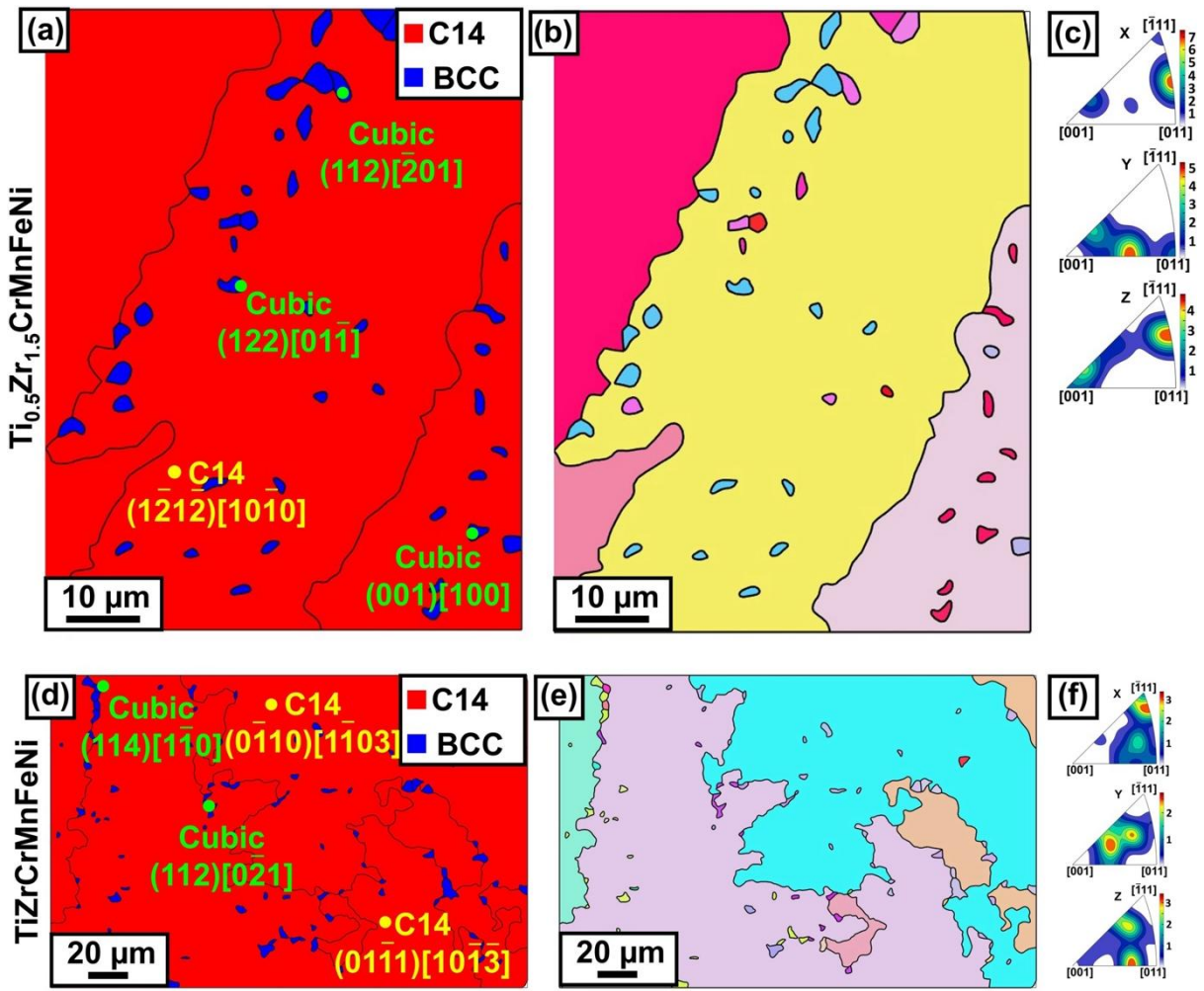

Figure 3. Random misorientations between BCC and C14 in two high-entropy alloys. Electron backscatter diffraction for (a-c) Ti$_{0.5}$Zr$_{1.5}$MnCrFeNi and (d-f) TiZrMnCrFeNi using (a, d) phase maps, (b, e) orientation maps and (c, f) inverse pole figures.

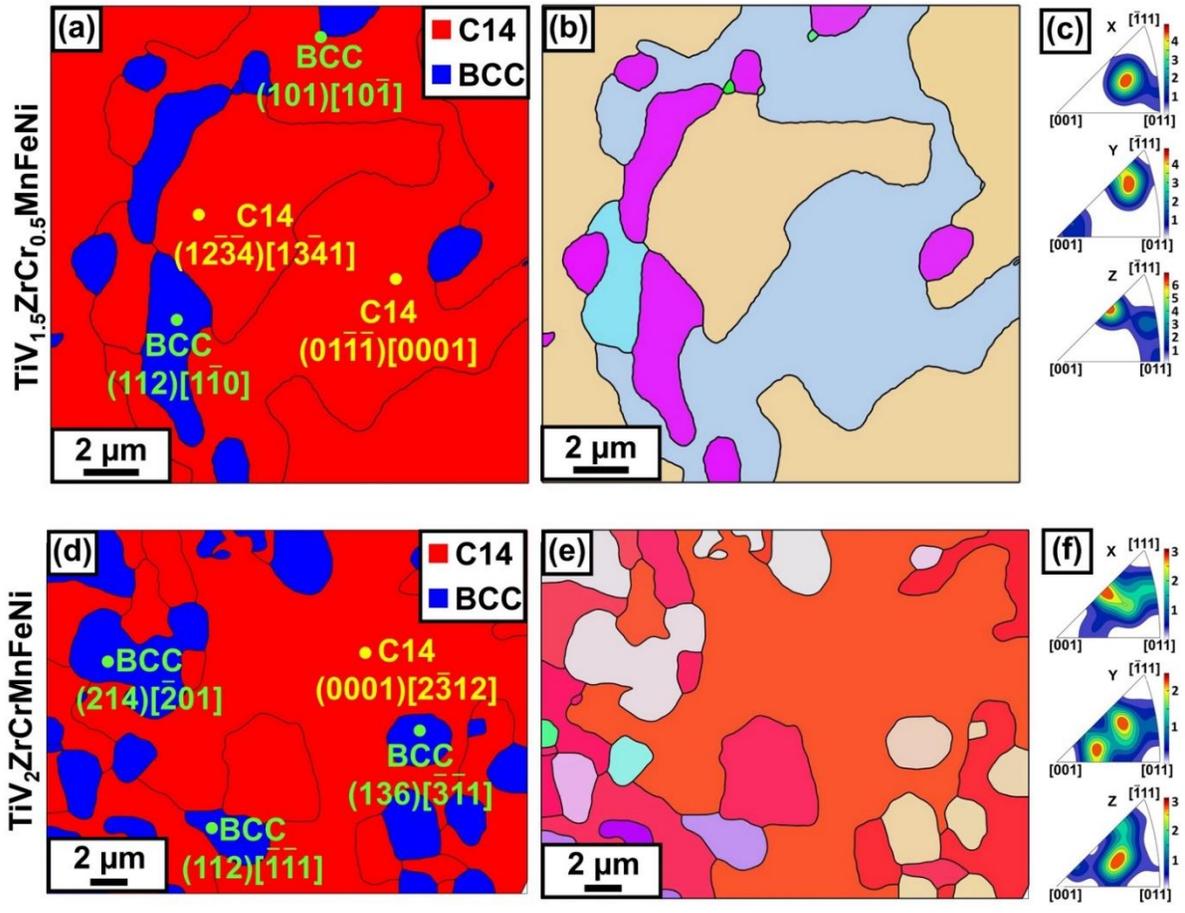

Figure 4. Random misorientations between BCC and C14 in two high-entropy alloys. Electron backscatter diffraction for (a-c) TiV$_{1.5}$ZrCr$_{0.5}$MnFeNi and (d-f) TiV$_2$ZrCrMnFeNi using (a, d) phase maps, (b, e) orientation maps and (c, f) inverse pole figures.

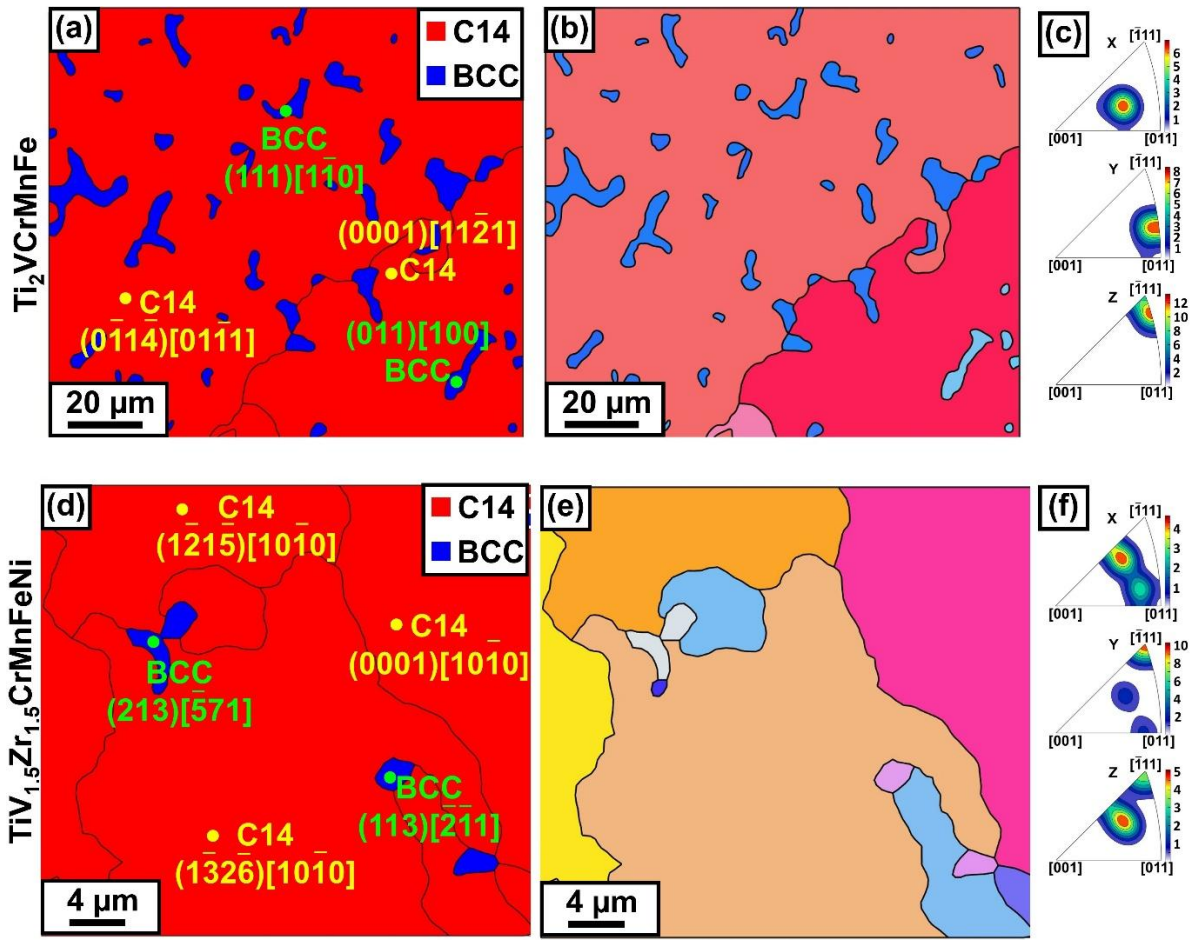

Figure 5. Uniform orientations between BCC and C14 in two high-entropy alloys. Electron back-scatter diffraction for (a-c) $Ti_2VCrMnFe$ and (d-f) $TiV_{1.5}Zr_{1.5}CrMnFeNi$ using (a, d) phase maps, (b, e) orientation maps and (c, f) inverse pole figures.

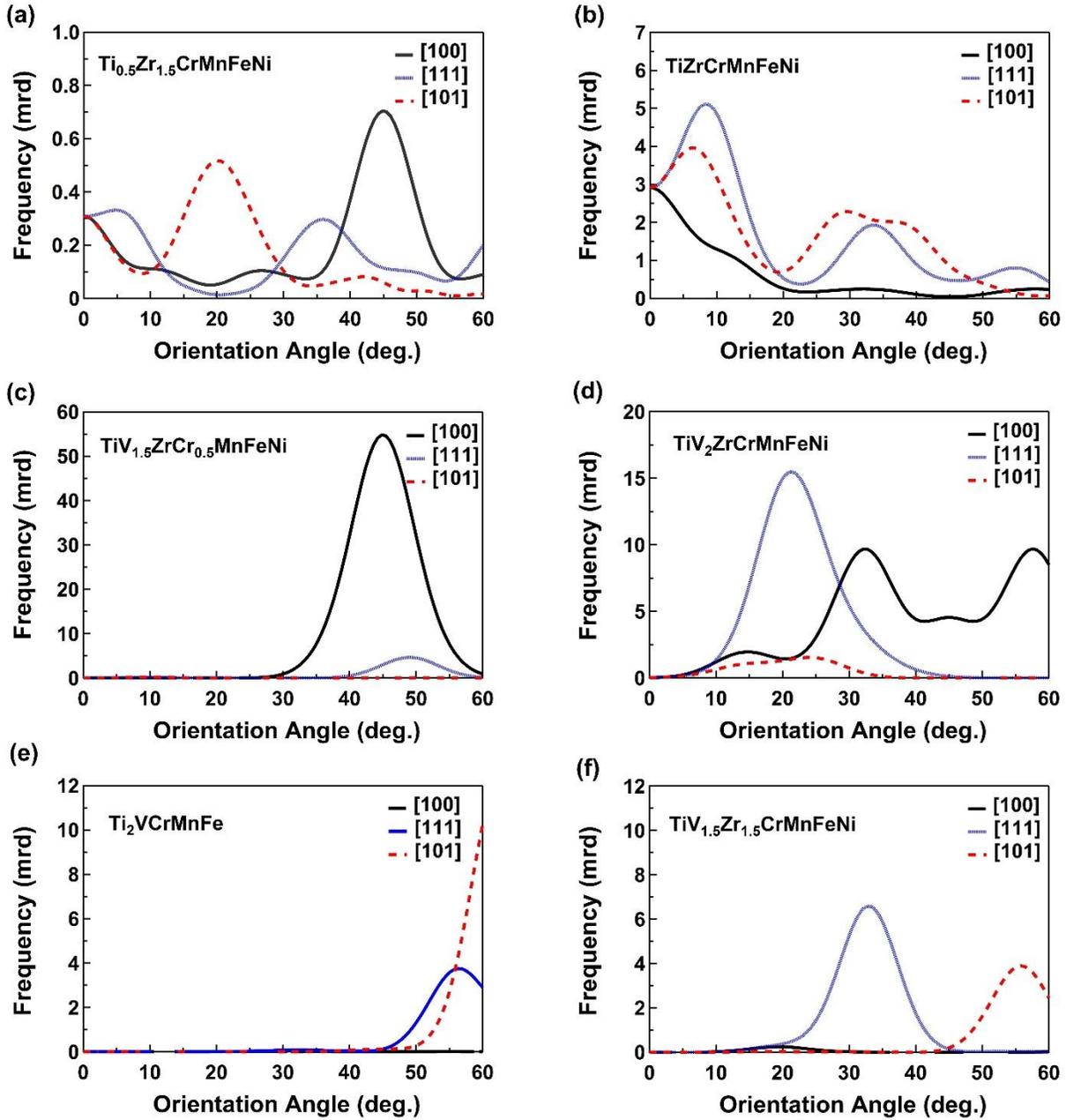

Figure 6. Random misorientation distribution of BCC in C14 matrix in most active high-entropy alloys. Frequency in multiple of random distribution (mrd) versus orientation angle of BCC with C14 Laves phase along [100], [111] and [101] fibers for high-entropy alloys (a) $Ti_{0.5}Zr_{1.5}CrMnFeNi$, (b) TiZrCrMnFeNi, (c) $TiV_{1.5}ZrCr_{0.5}MnFeNi$, (d) $TiV_2ZrCrMnFeNi$, (e) $Ti_2VCrMnFe$ and (f) $TiV_{1.5}Zr_{1.5}CrMnFeNi$.

### 3.4. Nanostructure of High-Entropy Alloys Examined by TEM

To confirm the coherency, high-resolution (HR) TEM images for one active HEA (TiV$_{1.5}$ZrCr$_{0.5}$MnFeNi) and one inactive HEA (Ti$_2$VCrMnFe) were taken. The HR-TEM image for the HEA TiV$_{1.5}$ZrCr$_{0.5}$MnFeNi is illustrated in Fig. 7(a) with the co-existence of both C14 Laves and BCC phases and an interphase boundary between them. The *d*-spacing of the C14 phase is 0.25 nm and the *d*-spacing of BCC is 0.16 nm. This confirms the interphase boundary as an incoherent boundary. Fig. 7 (b) shows HR-TEM for Ti$_2$VCrMnFe with an interphase boundary between them. A tendency to form a coherent interphase boundary with lower interfacial energy between the C14 and BCC phases is observed in Ti$_2$VCrMnFe. HR-TEM suggests a coherency relation as $(\bar{2}00)_{C14}$ // $(011)_{BCC}$ in this inactive material.

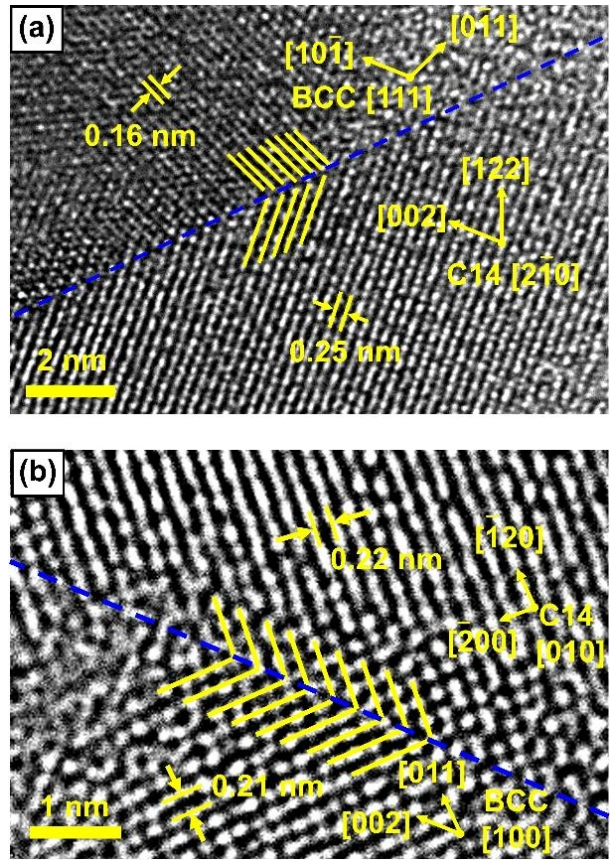

Figure 7. The presence of incoherent and coherent interphase boundaries between C14 Laves and BCC phases in two different high-entropy alloys. High-resolution TEM image of (a) incoherent interphase boundary in TiV$_{1.5}$ZrCr$_{0.5}$MnFeNi and (b) coherent interphase boundary in Ti$_2$VCrMnFe.

## 4. Discussion

One challenge for the widespread adoption of solid-state hydrogen storage is activation [20,21,22,45]. This issue of activation is not just restricted to metals and alloys but also to some HEAs such as Mg-Ti-V-Zr-Nb [46], TiVZrHfNb [47], TiVZrNbTa [48], TiZrHfMoNb [49], TiZrNbHfTa [50], TiZrFeNi [51] and ZrNbFeCo [52]. Having a material that can intrinsically absorb hydrogen without going through an additional step of activation will be a step closer towards the adoption of the solid-state hydrogen storage technique. One suggested method to solve this issue is by designing materials with interphase boundaries [45]. Interphase boundaries can function as hydrogen paths [20-24] as well as heterogeneous nucleation sites for hydrides [45,53-56]. In this study, it is observed that the activation of materials is not solely governed by interphase boundaries, but the crystallographic orientation of phases and their coherency also play a crucial role in determining their activity. Here, a question needs to be discussed. How does the coherency of phases affect the activation of materials, and what are the criteria for the materials to be active for hydrogen absorption by interphase boundaries?

Regarding this question, it should be considered that the orientation of grains does play a significant role in making the materials active. When grains are oriented in a form that atomic positions on either side of boundaries are correlated, then the boundary is termed a coherent boundary. Coherent interphase boundaries, due to their alignment of atoms, have lower free volume for hydrogen diffusion as compared to incoherent interphase boundaries [65]. Hydrogen flux through coherent interphase boundaries is likely similar to that of the grain interior [40]. Whereas, incoherent interphase boundaries, with their irregular atomic arrangements, offer increased free volume and enhance hydrogen diffusion [60,61,62,63,64]. Hydrogen flux through coherent interphases should be close to that of high-angle grain boundaries. Another issue is the lower boundary energy for coherent boundaries, which makes them stable, and this is inappropriate for the heterogeneous hydride nucleation [~~77~~,78,79]. Since some *d*-spacing differences are usually required to avoid coherency, this issue should be considered to make the material active by having *d*-spacing differences higher than 1%.

The *d*-spacing difference for HEA TiV$_2$ZrCrMnFeNi through XRD is less than 1%, but still, the HEA is an active HEA for hydrogen absorption. The amount of secondary BCC phase in this HEA is around 16 wt% and a large fraction of BCC is also visible in EBSD images in Fig.

4(d). Though some uniform orientation and coherency is observed in the EBSD orientation map in Fig. 4(e), incoherent interphase boundaries are unavoidable because of the large fraction of BCC. The IPF images in Fig. 4(f) show a spread of the BCC orientation, which indicates that the BCC phase exists with different alignments with the C14 phase in TiV$_2$ZrCrMnFeNi. This is also observed in MDF results presented in Fig. 6(d).

Considering the case of non-active HEA TiV$_{1.5}$Zr$_{1.5}$CrMnFeNi, some incoherency is observed between C14 and BCC in Fig. 5 (e). Moreover, MDF in Fig. 6(f) shows misorientation between C14 and BCC within the range of 30º to 40º, which is good for higher hydrogen diffusion according to the Frank-Bilby model [60,80,81]. However, HEA TiV$_{1.5}$Zr$_{1.5}$CrMnFeNi is inactive because of a very small amount of interphase boundaries. This contrasts with the HEA TiV$_2$ZrCrMnFeNi, in which, despite the presence of some coherent interphase boundaries and low *d*-spacing difference, full activity is achieved because of a larger fraction of interphase boundaries. Therefore, not only the incoherency but also an appropriate fraction of interphase boundaries is needed for activation of HEAs. Although a very large fraction of the second phase can activate the material, this negatively affects the storage capacity. It should be noted that incoherent interphase boundaries can lead to good cycling performance within long-term (i.e. 50-1000 cycles) [44,72], as their stability is better than normal high-angle grain boundaries produced by ball milling or severe plastic deformation [82].

## 5. Conclusion

This investigation highlights the significance of grain orientation and associated interphase boundary coherency on activation of high-entropy alloys (HEAs) for storing hydrogen at ambient temperature. Analyses show that in dual-phase C14+BCC HEAs, the presence of interphase boundaries is not the sole criterion for activation. The high coherency of interphase boundaries plays a negative role in the activity of HEAs for absorbing hydrogen, because coherent interphase boundaries are associated with minimal free volume and minimum interfacial energy and do not function as effective sites for hydrogen diffusion and hydride nucleation. Moreover, it is suggested that a higher fraction of incoherent interphases can activate the material more easily. These results recommend a solution for developing active high-entropy hydrogen storage alloys at ambient temperature by introducing incoherent boundaries. Future studies should quantitatively examine the correlation between chemical composition, interphase coherency and activation. Moreover, the generality of the concept of this study should be examined using other dual-phase HEAs, such as popular alloys with BCC + FCC phases. Additionally, translating the advantages of incoherent interphase boundaries into the practical sustainable hydrogen storage systems will require the development of scalable synthesis techniques such as induction melting.

**CRediT Authorship Contribution Statement**

All authors: Conceptualization, Methodology, Investigation, Validation, Writing – review & editing.

**Declaration of Competing Interest**

The authors declare no competing financial interests or personal relationships that can affect the research presented in this manuscript.

**Data availability**

All the data presented in this article are made available through the request to the corresponding author

**Acknowledgment**

The author Shivam Dangwal is grateful to the MEXT, Japan, for a scholarship. This research is partly funded by the ASPIRE project of the Japan Science and Technology Agency (JST) (JPMJAP2332).